\documentclass[aps,prd,preprint,superscriptaddress,tightenlines,nofootinbib,floatfix]{revtex4}

\usepackage{graphicx}
\usepackage{dcolumn}
\usepackage{bm}
\usepackage{multirow}

\begin{document}

\newcommand{\barb}{\bar{B}}
\newcommand{\bob}{\bar{B}^0}
\newcommand{\bo}{B^0}
\newcommand{\bp}{B^+}
\newcommand{\dkpipiz}{D^0 \to K^-\pi^+\pi^0}
\newcommand{\dkpi}{D^0 \to K^-\pi^+}
\newcommand{\dkrho}{D^0 \to K^- \rho^+}
\newcommand{\dobar}{\overline{D}^0}
\newcommand{\etal}{{\it et al.}}
\newcommand{\F}{$\cal F$}
\newcommand{\kpipiz}{K^-\pi^+\pi^0}
\newcommand{\krho}{K^- \rho^+}
\newcommand{\piz}{\pi^0}
\newcommand{\ra}{{\rightarrow}}
\newcommand{\Ufs}{\Upsilon(4S)}

\newcommand{\about}		{\mbox{$\sim$}}
\newcommand{\aerr}[3]   {\mbox{${{#1}^{+ #2}_{- #3}}$}}
\newcommand{\amp}    	{\mbox{${\cal A}$}}
\newcommand{\avcb}		{\mbox{$|V_{cb}|$}}
\newcommand{\avtb}		{\mbox{$|V_{tb}|$}}
\newcommand{\avtd}		{\mbox{$|V_{td}|$}}
\newcommand{\avts}		{\mbox{$|V_{ts}|$}}
\newcommand{\avub}		{\mbox{$|V_{ub}|$}}
\newcommand{\avud}		{\mbox{$|V_{ud}|$}}
\newcommand{\avus}		{\mbox{$|V_{us}|$}}
\newcommand{\BBbar}     {\mbox{$B\bar B$}}
\newcommand{\bbbar}     {\mbox{$B\bar B$}}
\newcommand{\bc}		{\mbox{$b\to c$}}
\newcommand{\bdb}		{\mbox{$\bar B^0_d $}}
\newcommand{\bd}		{\mbox{$B^0_d $}}
\newcommand{\berr}[2]   {\mbox{${{}^{+ #1}_{- #2}}$}}
\newcommand{\bmeson}	{\mbox{$B$}}
\newcommand{\bqb}		{\mbox{$\bar B^0_q $}}
\newcommand{\bq}		{\mbox{$B^0_q $}}
\newcommand{\branch}    {\mbox{${\cal B}$}}
\newcommand{\bsb}		{\mbox{$\bar B^0_s$}}
\newcommand{\bs}		{\mbox{$B^0_s$}}
\newcommand{\bu}		{\mbox{$b\to u$}}
\newcommand{\cerr}[4]   {\mbox{${{}^{+ #1}_{- #2}{}^{+ #3}_{- #4}}$}}
\newcommand{\chisq}		{\mbox{$\chi^2$}}
\newcommand{\cosB}		{\mbox{${\cos\theta_{\rm B}}$}}
\newcommand{\cossph}	{\mbox{$\cos \theta_{\rm sph}$}}
\newcommand{\costhr}	{\mbox{$\cos \theta_{\rm thr}$}}
\newcommand{\dedx}		{\mbox{$dE/dx$}}
\newcommand{\derr}[5]   {\mbox{${{#1}^{+ #2}_{- #3}{}^{+ #4}_{- #5}}$}}
\newcommand{\de}		{\mbox{$\Delta E$}}
\newcommand{\dzerobar}	{\mbox{$\overline {D^0}$}}
\newcommand{\dzero}		{\mbox{${D^0}$}}
\newcommand{\ebeam}		{\mbox{$E_{\rm beam}$}}
\newcommand{\eb}		{\mbox{$E_b$}}
\newcommand{\eeqq}		{\mbox{$e^+e^-\to\qqb$}}
\newcommand{\ee}		{\mbox{$e^+e^-$}}
\newcommand{\expt}		{\mbox{$_{\rm expt}$}}
\newcommand{\fbinv}		{\mbox{${\rm fb}^{-1}$}}
\newcommand{\fisher}    {\mbox{$x_{\cal F}$}}
\newcommand{\gev}		{\mbox{${\rm ~GeV}$}}
\newcommand{\hh}		{\mbox{$h^+h^-$}}
\newcommand{\hpm}		{\mbox{$h^\pm$}}
\newcommand{\implies}	{\mbox{${\Longrightarrow}$}}
\newcommand{\jimexp}[1]	{\mbox{${\rm e}^{{#1}}$}}
\newcommand{\kk}		{\mbox{KK}}
\newcommand{\Kpi}		{\mbox{$K\pi$}}
\newcommand{\kpi}		{\mbox{$\Kpi$}}
\newcommand{\kpz}		{\mbox{$K^+\pi^0$}}
\newcommand{\KP}		{\mbox{$K\pi$}}
\newcommand{\ksp}		{\mbox{$K^0_S\pi^+$}}
\newcommand{\ks}		{\mbox{$K^0_S$}}
\newcommand{\kz}		{\mbox{$K^0$}}
\newcommand{\kzb}		{\mbox{$\overline{K^0}$}}
\newcommand{\like}    	{\mbox{${\cal L}$}}
\newcommand{\Lp}		{\mbox{$\Lambda \bar p$}}
\newcommand{\lum}    	{\mbox{${\cal L}$}}
\newcommand{\mb}		{\mbox{$M_B$}}
\newcommand{\mev}		{\mbox{${\rm MeV}$}}
\newcommand{\micron}	{\mbox{$~\mu{\rm m}$}}
\newcommand{\mm}		{\mbox{$\mu^+\mu^-$}}
\newcommand{\model}		{\mbox{$_{\rm mod}$}}
\newcommand{\nbb}		{\mbox{$N_{B\bar B}$}}
\newcommand{\nbinv}		{\mbox{${\rm nb}^{-1}$}}
\newcommand{\pb}        {\mbox{$p_B$}}
\newcommand{\pbinv}		{\mbox{${\rm pb}^{-1}$}}
\newcommand{\pdf}		{\mbox{${PDF}$}}
\newcommand{\pdfs}		{\mbox{${PDF}{\rm s}$}}
\newcommand{\pidkpi}	{\mbox{$\Delta_{K\pi}$}}
\newcommand{\pidkp}		{\mbox{$\Delta_{Kp}$}}
\newcommand{\pid}		{\mbox{$\Delta_{PID}$}}
\newcommand{\pipi}		{\mbox{$\Pipi$}}
\newcommand{\Pipi}		{\mbox{$\pi\pi$}}
\newcommand{\power}[1]  {\mbox{${\times 10^{#1}}$}}
\newcommand{\ppz}		{\mbox{$\pi^+\pi^0$}}
\newcommand{\pvec}		{\mbox{$\vec{p}$}}
\newcommand{\pz}		{\mbox{$\pi^0$}}
\newcommand{\qqb}		{\mbox{$q\bar q$}}
\newcommand{\qq}		{\mbox{${q\bar q}$}}
\newcommand{\stat}		{\mbox{$_{\rm stat}$}}
\newcommand{\syst}		{\mbox{$_{\rm syst}$}}
\newcommand{\theo}		{\mbox{$_{\rm theo}$}}
\newcommand{\upsi}		{\mbox{$\Upsilon$({\rm 4S})}}
\newcommand{\vcb}		{\mbox{$V_{cb}$}}
\newcommand{\vcd}		{\mbox{$V_{cd}$}}
\newcommand{\vcs}		{\mbox{$V_{cs}$}}
\newcommand{\vtb}		{\mbox{$V_{tb}$}}
\newcommand{\vtd}		{\mbox{$V_{td}$}}
\newcommand{\vts}		{\mbox{$V_{ts}$}}
\newcommand{\vub}		{\mbox{$V_{ub}$}}
\newcommand{\vud}		{\mbox{$V_{ud}$}}
\newcommand{\vus}		{\mbox{$V_{us}$}}
\newcommand{\zhat}		{\mbox{$\hat{\bf z}$}}

\newcommand{\ppbar}		{\mbox{${p\bar p}$}}
\newcommand{\pL}		{\mbox{${p\bar\Lambda}$}}
\newcommand{\LL}		{\mbox{${\Lambda\bar\Lambda}$}}

\newcommand{\mpdf}    	{\mbox{${\cal M}$}}
\newcommand{\epdf}    	{\mbox{${\cal E}$}}
\newcommand{\fpdf}    	{\mbox{${\cal F}$}}
\newcommand{\cpdf}    	{\mbox{${\cal C}$}}
\newcommand{\dkmode}	{\mbox{${\mu}$}}
\newcommand{\contrib}	{\mbox{${\kappa}$}}
\newcommand{\mc}    	{\mbox{${}_{\dkmode\contrib}$}}
\newcommand{\nk}    	{\mbox{${n}_{\contrib}$}}

\newcommand{\G}		{\mbox{${G}$}}
\newcommand{\GG}	{\mbox{${\cal G}$}}
\newcommand{\ARG}	{\mbox{${A}$}}
\newcommand{\LIN}	{\mbox{${L}$}}
\newcommand{\BW}	{\mbox{${\cal R}$}}
\newcommand{\FI}	{\mbox{${F_0}$}}
\newcommand{\DG}	{\mbox{${a\G_1+b\G_2}$}}
\newcommand{\GGG}	{\mbox{${a\G+b\GG}$}}
\newcommand{\DGG}	{\mbox{${a\GG_1+b\GG_2}$}}

\newcommand{\sba} 	{\mbox{${S/B}$}}
\newcommand{\rfw}   {\mbox{$R_2$}}
\newcommand{\effmc}     {\mbox{${\epsilon_{\rm MC}}$}}
\newcommand{\effdata}     {\mbox{${\epsilon_{\rm DATA}}$}}

\newcommand{\RR}	{\mbox{${\cal R}$}}


\preprint{CLNS 03/1816}       
\preprint{CLEO 03-03}         

\title{Measurements of Charmless Hadronic Two-Body B Meson Decays 
and the Ratio $\branch(B\to DK)/\branch(B\to D\pi)$}


\author{A.~Bornheim}
\author{E.~Lipeles}
\author{S.~P.~Pappas}
\author{A.~Shapiro}
\author{W.~M.~Sun}
\author{A.~J.~Weinstein}
\affiliation{California Institute of Technology, Pasadena, California 91125}
\author{R.~A.~Briere}
\author{G.~P.~Chen}
\author{T.~Ferguson}
\author{G.~Tatishvili}
\author{H.~Vogel}
\affiliation{Carnegie Mellon University, Pittsburgh, Pennsylvania 15213}
% ...
\collaboration{CLEO Collaboration} 
\noaffiliation



\date{\today}

%
%

\begin{abstract} 
We present final measurements of thirteen charmless hadronic \bmeson\ decay modes from the
CLEO experiment. The decay modes include the ten \pipi, \Kpi, and \kk\
final states and new limits on dibaryonic final states, \ppbar, \pL, and
\LL, as well as a new determination of the ratio $\branch(B\to
DK)/\branch(B\to D\pi)$. The results are based on the full CLEO II and
CLEO III data samples totalling 15.3 \fbinv\ at the $\Upsilon(4S)$, and
supercede previously published results.

\end{abstract}
\maketitle

%
\section{Introduction}
%

Charmless decays of \bmeson\ mesons may proceed by $b\to u$, $b\to s$,
or $b\to d$ transitions.  The latter two mechanisms require flavor
changing neutral currents which are not present at tree level in the
Standard Model, and therefore must occur through higher order processes
such as the penguin mechanism.  Such processes involve loops, which can
open the window for particles and physics outside the Standard Model.
Even in the absence of such new physics, interference among competing
amplitudes for a given decay mode can be exploited to measure CKM
phases. There is a significant body of literature\cite{fleischeretal} on
the use of the branching ratio $\branch(B\to K^-\pi^+)$ and other
charmless modes to determine or constrain the CKM angle $\gamma$, the
phase of \vub\ in conventional representations of the CKM matrix.
Compared to the methods of extracting $\gamma$ that are based the $B\to DK$
decay modes\cite{DK}, these approaches based on charmless decay modes are less
clean theoretically, but more promising experimentally because the
event yields are significantly higher. 

Recent work on two-body charmless decay modes suggests that the
unitarity triangle may be constructable entirely from charmless modes,
without recourse to the traditional constraints involving \bmeson\
mixing measurements, CP asymmetry in $B\to J/\psi \ks$, or CP violation
in kaon decays. The charmless modes therefore offer an independent
approach to probe CP violating effects in heavy quark decay. Significant
disagreement between these two approaches, if found in experimental
results, would directly challenge the Standard Model and its fundamental
statement that all CP violating phenomena stem from a single phase in
the CKM matrix. Early results based on current data are already
available\cite{neubertnew}, and do indicate a degree of inconsistency. In
this paper we present new experimental data on charmless modes and note
that these data enhance rather than ameliorate the discrepancy.

CLEO has previously published several papers\cite{previous} reporting
measurements of charmless hadronic \bmeson\ meson decay modes, including
searches for charmless baryonic final states, with the data of the CLEO
II experiment.  Here we report corresponding measurements in the new
CLEO III data with results for three $\pi\pi$ modes, $B\to \pi^+\pi^-,~
\pi^+\pi^0,~\pz\pz$, four $K\pi$ modes, $B\to K^+\pi^-,~K^+\pi^0,~
\kz\pi^+,~\kz\pi^0$, three $K\overline{K}$ modes, $B\to K^+K^-,
~\kz K^-,~\kz\kzb$, and three dibaryonic modes, $B\to
\ppbar,~\pL,~\LL$. We also merge CLEO II and CLEO III results to
determine a final measurement for each mode based on the full CLEO data
set, which hereby supercedes our previous publications.  Recent
measurements from BABAR and Belle are in excellent agreement with
ours\cite{babarandbelle}. We also report a new measurement of the ratio
of branching ratios, $\branch(B\to D^0 K^-)/\branch(B\to D^0\pi^-)$.  

Here and throughout this paper charge conjugate modes are implied. We
also make use of the notation $h^\pm$ to represent a charged hadron that may be
either a kaon or pion.

%
\section{The CLEO Detector and Datasets}
%
CLEO is a general purpose solenoidal magnet detector operating at the
Cornell Electron Storage Ring (CESR). The latter is a symmetric-energy
storage ring tuned for the data sets discussed here to provide center of
mass energies near the \upsi. At $\sqrt{s}=M_{\upsi}$ the hadronic cross
section is approximately 4 nb, with 1 nb of $\ee\to\upsi\to\bbbar$ and 3
nb of four-flavor continuum \eeqq.  In the CLEO III running period, July
2000 through June 2001, we obtained an integrated luminosity of 6.18
\fbinv\ at the \upsi\ and 2.24 \fbinv\ off-resonance, {\it i.e.,} just
below the \bbbar\ threshold.  The off-resonance data are used for
background determinations.  The on-resonance data corresponds to $\nbb =
(5.73\pm 0.47)\power{6}$ \upsi\ decays. The corresponding numbers for
the CLEO II running period (1990-1999) are 9.13 \fbinv\ (($9.66\pm
0.19)\power{6} $\upsi\ decays) and 4.35 \fbinv.  Differences in the \nbb\
yield per unit integrated luminosity reflect differences in run
conditions.

The CLEO III detector\cite{cleo3det} differs from the CLEO II
detector\cite{cleo2det} most notably in the inclusion of a ring-imaging
Cherenkov device (RICH)\cite{rich} which provides particle
identification at all momenta above the Cherenkov threshold. Even at the
highest momenta relevant for \bmeson\ physics, about 2.8 \gev, the RICH
separates kaons and pions by 2.3 standard deviations. Measurements of
specific ionization (\dedx) in the drift chamber provide an additional
2.0 standard deviation separation at the highest momenta. Charged
particle tracking is done by the 47-layer drift chamber and a four-layer
silicon tracker which reside in a 1.5T solenoidal magnetic field and
provide momentum resolution described by $(\sigma_p/p)^2 = A^2 + B^2
p^2$ with $A \approx 0.005$ and $B\approx 0.001~\gev^{-1}$. The
absolute momentum calibration is confirmed by comparing the invariant
mass of standard decays $J/\psi \to \mu^+\mu^-$, $D^0 \to K^-\pi^+$ with
PDG values\cite{pdg}. Photons are detected using a 7800-crystal CsI(Tl)
electromagnetic
calorimeter
which is unchanged between CLEO II and CLEO III.

%
\section{Elements of the analysis}
%
The \upsi\ is produced at rest in the lab frame and decays with low $Q$
value to a pair of \bmeson\ mesons that travel non-relativistically,
with $\pb \sim 300~\mev$.  In this analysis we assume equal rates of
$B^+B^-$ and $B^0 \bar{B^0}$ production\cite{fplusminuscleo}. All decay
modes studied in this paper are two-body or quasi-two-body modes. Apart
from the modest $\pm 150~\mev$ doppler shifts resulting from the motion
of the \bmeson\ mesons in the lab frame, the daughter particles are
nearly monochromatic, and, up to resolution smearing, jointly carry the
full beam energy \eb\ and have invariant mass equal to the \bmeson\ mass
\mb. The approximate monochromaticity of the daughters simplifies
particle identification and energy resolution, and helps keep the
associated systematic errors low. The other \bmeson\ in the event decays
into, on average, five charged and five neutral particles, distributed
uniformly in the detector acceptance. The principal background to the
analysis comes from the non-$b$ hadronic data, \eeqq, with $\qq = u\bar
u, ~d\bar d, ~s\bar s, ~{\rm and}~ c\bar c.$ High momentum, back-to-back
particles are typical in such events, and some have invariant masses and
total energies close to or in the signal region of the \bmeson\ events. 
Fortunately, distinctive event topologies separate most of these
background events from the signal.  

This analysis has two principal parts: (a) the application of hard 
selection criteria to obtain
signal-like events, based on kinematics, event topology, and
particle identification; (b) the application of an unbinned extended
maximum likelihood fit to the surviving event ensembles to extract the
yields of signal and background(s) for each mode. The likelihood fit
allows us to make maximum use of available information, while avoiding
efficiency losses that further selection criteria would entail.

For the purposes of reconstruction, the CLEO III dataset reported on
here is divided into two subsets of roughly equal integrated luminosity,
which we will call Set A and Set B. The distinction has ultimately no
significant effect on the results, but because of changes in event
reconstruction algorithms between the two sets, there are slight
differences in resolutions and efficiencies -- mostly affecting modes
with charged particles -- that we treat separately until the final CLEO
III results are reassembled at the end. We provide in Table
\ref{tab:oldnew} some informative comparisons between Set A and Set B.

\begin{table}
\begin{center}
\caption{Features of Set A and Set B.}
\smallskip
\begin{tabular}{lcc}
\hline
Quantity & ~~~~Set A & ~~~~Set B\cr 
\hline\hline
Fraction of total \nbb\     & 55\%    & 45\%  \cr
Track Resolution            &         &       \cr
\hfil{`A' Coefficient}      & 0.0055  & 0.0044\cr
\hfil{`B' Coefficient ($\gev^{-1}$)}      & 0.0011  & 0.0010\cr
$K^+\pi^-$ Mode             &         &       \cr
\hfil{$\sigma_{\mb}$ (\mev)}  & 2.7     & 2.7   \cr
\hfil{$\sigma_{\de}$(\mev)}   & 22      & 19    \cr
\hfil{Efficiency}           & 38\%    & 45\%  \cr
$K^+\pi^0$ Mode             &         &       \cr
\hfil{$\sigma_{\mb}$ (\mev)}  & 3.1     & 3.1   \cr
\hfil{$\sigma_{\de}$(\mev)}   & 31      & 31    \cr
\hfil{Efficiency}           & 33\%    & 35\%  \cr
$\pi^0\pi^0$ Mode\footnote{Resolutions are given as average of low-side and high-side half-resolutions.}           &         &       \cr
\hfil{$\sigma_{\mb}$ (\mev)}  & 3.6     & 3.6  \cr
\hfil{$\sigma_{\de}$(\mev)}   & 43      & 43    \cr
\hfil{Efficiency}           & 22\%    & 22\%  \cr
\hline
\end{tabular}
\label{tab:oldnew}
\end{center}
\end{table}

Charged track and photon candidates are required to satisfy loose
quality requirements which reject poorly determined candidates while
retaining high efficiency for real tracks and showers. \ks\ candidates
are selected from pairs of charged tracks forming well-measured
displaced vertices with a $\pi^+\pi^-$ invariant mass within three
standard deviations of the nominal \ks\ mass. In addition the vertex
must satisfy $|r_{VTX}| > 5$mm in the transverse plane, and
${\vec{p}_{\ks}}\cdot {\vec{r}_{VTX}} > 0 $. The $\ks\to\pz\pz$ mode is
not used.  $\Lambda$ candidates consist of $p\pi$ pairs with invariant
mass within three standard deviations  of the nominal $\Lambda$ mass.
Pairs of photons with an invariant mass within 2.5 standard deviations
of the nominal $\pi^0$ mass are kinematically fit with the mass
constrained to the nominal $\pi^0$ mass.

%
\subsection{General Event Selection}
%
Candidates for rare $B$ decay events are selected for further analysis
on the basis of two kinematic variables and one event-shape variable.
For each candidate, we construct the beam constrained $B$ candidate mass $M_B =
\sqrt{(E_b^2-{\vec{p}}^{~2})}$ where $E_b$ is the beam energy, and
$\vec{p}$ is the momentum of the candidate computed from the vector sum
of the daughter momenta. For real \bmeson\ mesons $|\vec{p}| \ll E_b$
and the width of this distribution is dominated by the $\sim 2.5~\mev$
intrinsic beam energy spread. The beam energy is determined run by run
from CESR lattice information, and slight corrections are applied
afterward to ensure that the observed $B^-$ mass in $B^-\to D^0\pi^-$
events matches the accepted value\cite{pdg}. In addition we compute the
energy balance variable $\Delta E$ = $E - E_b$ where $E$ is the sum of
the daughter energies.  The width of this distribution is about 20 \mev\
in all charged modes, as determined by the momentum resolution of the
tracking systems, and is about 40 \mev\ in modes involving neutral pions.

Any candidate with $|\Delta E|<400$ MeV and $M_B > 5.2$ GeV is
kept. An additional requirement on \cossph, the cosine of the angle
between the sphericity axis of the candidate and the sphericity axis of
the rest of the event\cite{previous}, is used to reject the dominant
\eeqq\ background. All candidates must satisfy the requirement
$|\cossph|<0.8$, which rejects approximately 80\% of the background
while retaining nearly 80\% of the signal.

%
\subsection{Particle Identification Requirements (CLEO III)}
%
In the case of a candidate mode involving one or more charged pions or
kaons, such as $B\to \Kpi$ or $B\to \pi \pz$, each charged track must be
positively identified as $K$ or $\pi$.   The pattern of Cherenkov photon
hits in the RICH detector is fit to both a kaon and pion hypothesis,
each with its own likelihood $\like_K$ and $\like_\pi$.  The mean number
of photon hits entering the fit is twelve, and we require a minimum of
four.  Calibrated \dedx\ information from the drift chamber is used to
compute a \chisq\ for kaon and pion hypotheses.  The RICH and \dedx\
results are combined to form an effective \chisq\ difference, 
\begin{equation} 
\pidkpi = -2\ln\like_K + 2\ln\like_\pi + \chisq_K - \chisq_\pi. 
\label{eq:pidkpi-def}
\end{equation} 
Kaons are identified by $\pidkpi < \delta_K$ and pions by $-\pidkpi <
\delta_\pi$, with values of $\delta_K$ and $\delta_\pi$ chosen to yield
$(90\pm 3)$\% efficiency as determined in an independent study of tagged
kaons and pions obtained from the decay $D^{*+}\to\pi^+D^0~(D^0\to
K^-\pi^+)$.  With this choice of $\delta_K$ and $\delta_\pi$, the
misidentification rate for kaons faking pions (pions faking kaons) is
11\%(8\%) at momenta around 2.6 \gev. 

For candidate modes involving protons, positive proton identification is
required.  \dedx\ does not distinguish well between protons and kaons at
the $\sim 2.5~\gev$ momenta of interest, however, so the proton-kaon
separation is achieved with a discriminant based only on RICH
information: $\pidkp = -2\ln\like_ p+ 2\ln\like_K < \delta_p$. In this case
$\delta_p$ is chosen to yield proton (antiproton) identification
efficiency of $76\pm 1 (72\pm 1)$\% with a kaon fake rate of 1\%, as
determined in an independent study using tagged kaons from the $D^{*+}$
sample as above, and protons from $\Lambda \to p\pi$ decays.

%
\subsection{Event Selection for CLEO II modes}
%
We present three results for which we also analyzed the full CLEO II
data set, namely $B\rightarrow \kz\kzb$, $B\rightarrow
\Lambda\overline{\Lambda}$, and $B\rightarrow p \overline{\Lambda}$. The
$\ks$ selection required that the $\ks$ vertex is separated from the
beam spot by more that 3 sigma (5.5 sigma for CLEO II.V for which the
innermost drift chamber was replaced with a 3 layer silicon vertex
detector). The candidate mass must lie within 10 MeV of the nominal
$\ks$ mass. We require that the $\ks$ flight direction points to
within 3 sigma of the beamspot .

The protons in the $\pL$ final state must be compatible
within 3 sigma with a proton \dedx\ hypothesis and incompatible 
with both the electron (calculated from calorimeter information) and muon
(calculated from muon chamber information) hypotheses. We require that 
the $\Lambda$ candidate mass lie within 10 MeV of the nominal $\Lambda$
mass, the vertex be at least 5 mm removed radially from the beam spot,
and the $\chi^2$ of the vertex fit be less than nine. 
There is no particle identification applied to daughter particles of the
$\Lambda$ decays.

%
\section{Analysis Variables}
%
Events which meet all the requirements described in the preceding
paragraphs are now used in a likelihood fit to extract signal yield.
We characterize each candidate event by four variables: the mass and
energy variables introduced above, \mb\ and \de, the flight direction
of the candidate \bmeson, and a Fisher discriminant\cite{Fisher}. 

The flight direction is given by $\cosB = \hat{\bf p}\cdot \zhat$
where ${\bf p}$ is the vector sum of the daughter momenta and \zhat\ is
the beam axis. Since the vector \upsi\ is produced in \ee\ annihilation
it has a polarization $J_z=\pm 1$, and the subsequent
flight direction of the pseudoscalar \bmeson\ mesons 
is distributed as $|Y^{\pm 1}_1(\theta,\phi)|^2\sim\sin^2\theta = 1-\cos^2\theta$. 
Background events are flat in this variable.

The Fisher discriminant is used to refine the separation of signal and
\eeqq\ background that is initially addressed by the hard cut on
\cossph\ in the general event selection. The Fisher discriminant,
\fisher, is a linear combination of fourteen variables, with
coefficients chosen to maximize the separation of signal and background
events.  The optimization procedure uses Monte Carlo events for the
signal and off-resonance and (\mb,\de) sideband data events for the
background. As in previous CLEO publications\cite{previous} the
component variables include the direction of the thrust
axis of the candidate with respect to the beam axis,
\costhr, and the nine conical bins of a ``Virtual Calorimeter" whose axis
is aligned with the candidate thrust axis. A fuller description of the
Virtual Calorimeter is available in a previous 
publication\cite{bigrareb}. Note that \costhr\ and \cosB\ are quite
different quantities.  For two body decay $B\to XY$, one has simple
closed form expressions:  $\cosB = \hat{\bf p}\cdot \zhat$
with ${\bf p}\equiv{\bf p}_X+{\bf p}_Y$, whereas 
$\costhr  = \hat{\bf q}\cdot \zhat$ with ${\bf q}\equiv{\bf p}_X-{\bf p}_Y$.

In addition, we take advantage of the high quality particle
identification in CLEO III to augment the Fisher discriminant with
information on the presence of electrons, muons, protons, and kaons in
the event. The momentum of the highest momentum electron, muon, kaon,
and proton are used as inputs to the Fisher discriminant.  For these
purposes we need only rudimentary particle identification criteria. If
any of the possible particle type 
hypotheses has no corresponding track identified (which is very often the
case), a value zero is used as the input to the Fisher discriminant. 

The Fisher variable thus defined provides discrimination between 
charmless \bmeson\ decay signal modes and \qq\ background at a level
equivalent to two gaussian distributions separated by 1.4$\sigma$, 
and is independent of the details of the signal mode for all the modes
studied here.  

%
\section{Likelihood Fit}
%
%
\subsection{Fit Components}
%
With the four analysis variables \mb, \de, \fisher, and \cosB, we
characterize each event in terms of normalized probability distribution
functions (\pdfs): $\mpdf\mc(\mb)$, $\epdf\mc(\de)$,
$\fpdf\mc(\fisher)$, and $\cpdf\mc(\cosB)$. The thirteen different
charmless decay modes to be fit
will in general have contributions from (a) signal, (b) \qq\ background,
and (c) cross-feed from other \bmeson\ modes. Subscripts \dkmode\ and
\contrib\ identify the particular decay mode (\dkmode) and the type of
contribution (\contrib). The probability that a given event characterized
by (\mb, \de, \fisher, \cosB) is an event of component type \contrib\ of
decay mode \dkmode\ is then given by the product of \pdfs 

\begin{equation}
P\mc=
\mpdf\mc(\mb)\epdf\mc(\de)\fpdf\mc(\fisher)\cpdf\mc(\cosB).
\label{eq:evtprob}
\end{equation}

We determine the yields \nk\ of signal, \qq\ background, 
and cross-feed background in decay mode \dkmode\ by maximizing
the extended likelihood function with respect to the yields \nk:

\begin{equation}
\like_{\dkmode} (n_{sig}, n_{\qq}, n_{xfeed})= {\exp(-{ \sum_{\contrib} {\nk}})
\prod_{\rm events}{\left( \sum_{\contrib} {\nk P\mc}\right)}.}
\label{eq:likelihood}
\end{equation}

The \qq\ background is the dominant background source in all cases, and
in only five of the fifteen modes do we need to include any cross-feed
backgrounds.  Four of these are due to the $\sim 10\%$ $K/\pi$
misidentification probability.  In fitting $B\to\pi^+\pi^-$ and
$B\to\pi^+\pi^0$ we include components for $B\to K^+\pi^-$ and $B\to
K^+\pi^0$, respectively;  in $B\to K^+ K^-$ we include a component for
$B\to K^+\pi^-$; and in $B\to D^0 K^-$ we include a component for $B\to
D^0\pi^-$. Although the cross-feed backgrounds arise from mistaken
particle identification, they are still distinguishable from the signal
through $\epdf(\de)$, which is shifted by about 50 \mev\ relative to the
signal \pdf. The cross-feed fits are only for background removal and the
yields are not used in any other signal determination.

The fifth mode requiring a cross-feed component is $B^0\to\pz\pz$. In
this case a small contribution from $B^+\to\rho^+\pi^0$ arises when the
charged pion has very little momentum in the lab frame. Although the
missing particle also shifts \de\ by at least one pion mass, resolution
smearing leaves a small tail in the signal region. The treatment here is
the same as in our previous publication on
$B^0\to\pz\pz$\cite{previous}. We note also that potential feedthrough
of $B^+\to\rho^+\pi^0$ into $B^+\to \pi^+\pz$ is smaller than in the
\pz\pz\ case because the low-side resolution smearing is less for the
$\pi^+\pz$ mode, and because the ratio $\branch(B^+\to
\pi^+\pz)/\branch(B^+\to\rho^+\pi^0)$ is larger than $\branch(B^+\to
\pz\pz)/\branch(B^+\to\rho^+\pi^0)$. Monte Carlo studies confirm these
observations and we therefore do not include this term in the $\pi^+\pz$
fit.

%
\subsection{\pdfs}\label{sec:pdfs}
%
We parametrize the \pdfs\ with various functions and combinations of
functions which are listed below. In each case the parameters of the
function are determined from a fit to signal Monte Carlo event samples
for the signal component and cross-feed component (if there is one), and
from a fit to off-resonance data for the \qq\ background component. 
These parameters are then fixed for all subsequent fitting procedures so
the only free variables in the likelihood function Eq.
\ref{eq:likelihood} are the signal and background yields.  There is of
course underlying uncertainty in the parameter values which fix the
\pdf\ shapes, but this uncertainty is systematic in nature and will be
discussed later in section \ref{sec:syst}. All functions are normalized
to unit area over the accepted range of the free variable.

\begin{itemize}

\item Gaussian (\G): 
used for \mb\ and \de\ signal component \pdfs\ that do not involve
neutral pions.
The parameters are the mean and width.   

\item Asymmetric Gaussian (\GG): 
used for \mb\ and \de\ in modes where neutral pions appear. Fluctuations in the
measured \pz\ energy are intrinsically asymmetric -- with a longer tail
on the low energy side -- because of energy leakage out the back of the
CsI crytals in the electromagnetic calorimeter. The parameters are the
mean and separate left and right widths.

\item Linear (\LIN): 
used for \qq\ backgrounds in \de.  The free parameter is the slope.

\item ARGUS (\ARG): 
used to characterize the \mb\ shape of \qq\
backgrounds\cite{argus}. $\ARG(\mb) =
\sqrt{1-\mb^2/E_b^2}\exp(-\lambda(1-\mb^2/E_b^2))$. The parameter
$\lambda$ governs the turn-over of the shape and the slope at low values
of \mb. The beam energy $E_b$ determines the endpoint of the \qq\ \mb\
spectrum.  Over the course of CLEO III data taking this end point
clusters around several close but not identical values.  In practice we
form a sum of three ARGUS functions with different endpoint values,
weighted by the corresponding integrated luminosities. In addition, to
account for run-to-run beam energy variation, we
convolve each ARGUS function with a Gaussian of width $\sigma \sim
0.7~\mev$ in $E_b$.

\item Breit-Wigner (\BW): 
used in Fisher parametrizations to describe
non-Gaussian tails. Parameters are mean and width.

\item Fisher (\FI): 
a linear combination of functions used to
characterize the \qq\ background Fisher shape.  
It is primarily  an asymmetric Gaussian (87\% of the area), but includes an
additional Breit-Wigner (9\%) with the same mean, and a small symmetric Gaussian (4\%). 
\end{itemize}

Table \ref{tab:allfits} lists the \pdfs\ used for each fit component of
each mode.  The fourth fit variable, \cosB\ is in all cases taken to have
the functional form $1-\cos^2\theta_{\rm B}$ for signal 
and cross-feed components, and flat for \qq\ background.

\begin{table}
\begin{center}
\caption{Functional forms used in likelihood fits. See text for
discussion of terms. Linear combinations are indicated by
coefficients $a$ and $b$.}
\smallskip
\begin{tabular}{llccc}
\hline
Mode & Fit Component & \mpdf(\mb) & \epdf(\de) & \fpdf(\fisher)  \cr 
\hline\hline 
$\pi^+\pi^-$ & Signal    & \G\   & \G\   & \GG\ \cr
              & \qq\ \footnote{The \qq\ \pdfs\ are the same for all modes. For brevity we omit this line in subsequent entries.}     & \ARG\ & \LIN\  & \FI \cr
             & Cross-feed & \G\   & \G\   & \GG\ \cr
$\pi^+\pi^0$ & Signal    & \GG\   & \DGG\   & \GG\ \cr
             & Cross-feed & \G\   & \G\   & \GG\ \cr
$\pi^0\pi^0$ & Signal    & \GG\   & \DGG\   & \GG\ \cr
             & Cross-feed & \GG\   & \GG\   & \GG\ \cr
\hline
$K^+\pi^-$   & Signal    & \G\   & \G\   & \GG\ \cr
$\kz\pi^+$   & Signal    & \GG\   & \DG   & \GG\ \cr
$K^+\pi^0$   & Signal    & \GG\   & \DGG   & \GG\ \cr
$\kz\pi^0$   & Signal    & \G\   & \G\   & \GG\ \cr
\hline
$K^+K^-$     & Signal    & \G\   & \G\   & \GG\ \cr
             & Cross-feed & \G\   & \G\   & \GG\ \cr
$\kz K^-$     & Signal    & \GG\   & \DG\   & \GG\ \cr
$\kz\kzb$     & Signal    & \GG\   & \DG\   & \GG\ \cr
\hline
$p\bar p$    & Signal    & \GG\   & \G\   & \DG\ \cr
$\pL$        & Signal    & \GG\   & \G\   & \DG\ \cr
$\LL$        & Signal    & \GG\   & \G\   & \DG\footnote{Set A includes \BW.} \cr
\hline
$D^0\pi^-$   & Signal    & \GG\   & \G\   & \GG\ \cr
$D^0 K^-$    & Signal    & \GG\   & \G\   & \GG\ \cr
             & Cross-feed & \GG\   & \G\   & \GG\ \cr
\hline
\end{tabular}
\label{tab:allfits}
\end{center}
\end{table}

%
\subsection{Fit Results}
%
Table \ref{tab:results} shows the results of the fits to the CLEO III
data.  All errors shown are statistical only.  The apparently
large yields of
\qq\ background reflect the large 
background-normalizing sidebands in \mb\ and \de\ and are
not indicative of $S/B$. Typically $S/B\sim 1$ for
the observed modes.

\begin{table}
\begin{center}
\caption{Results of likelihood fits: raw event yields with statistical errors.
A dash in the last column means no cross-feed term was used in the fit.}
\smallskip
\begin{tabular}{lccccc}
\hline
 Mode& Set  & Eff (\%)  & Signal & \qq\ Bkg & cross-feed \cr 
\hline\hline 
$\pi^+\pi^-$ & A:  &  39.0 &	$7.8^{+5.5}_{-4.5}$ &	1750$\pm$42 & $3.9^{+4.6}_{-3.7}$ \\
             & B:  &  45.3 &	$4.3^{+4.1}_{-3.1}$ &	1955$\pm$44 & $2.8^{+3.4}_{-2.3}$ \\
	         
$\pi^+\pi^0$ & A:  &  34.9 &	$2.8^{+3.3}_{-1.9}$ &	1158$\pm$34 &	$9.3\pm 7.0$ \\
	         & B:  &  37.5 &	5.7$\pm$5.9 &		1139$\pm$34 &	$0.0\pm 3.5$ \\

$\pi^0\pi^0$ & A:  &  22.1 &	$2.2^{+2.5}_{-1.5}$ &	134$\pm$12 & $3.6^{+2.3}_{-3.1}$ \\
	         & B:  &  22.4 &	$0.4^{+2.7}_{-3.4}$ &	211$\pm$15 & $0.5^{+1.7}_{-2.6}$ \\
\hline
$K^+\pi^-$ & A:  &  37.9 & 	$28.1^{+6.8}_{-6.0}$ &	1779$\pm$42 &	$-$ \\
	       & B:  &  45.3 &	$19.1^{+5.3}_{-4.6}$ &	1848$\pm$43 &	$-$ \\

$\kz\pi^+$ & A:  &  12.3 &	$12.1^{+4.4}_{-3.7}$ &	398$\pm$20 &	$-$ \\
	       & B:  &  12.8 &	$2.9^{+1.8}_{-2.7}$ &	395$\pm$20 &	$-$ \\

$K^+\pi^0$ & A:  &  32.6 &	$16.7^{+6.2}_{-5.3}$ &	735$\pm$27 &	$-$ \\
	       & B:  &  35.3 &	$10.8^{+5.1}_{-4.1}$ &	780$\pm$28 &	$-$ \\

$\kz\pi^0$ & A:  &  9.6 &	$3.5^{+2.8}_{-1.9}$ &	154$\pm$13 &	$-$ \\
	       & B:  &  10.5 &	$2.9^{+2.4}_{-1.6}$ &	132$\pm$12 &	$-$ \\
\hline
$K^+K^-$ & A:  &  35.2 &	$2.3^{+3.7}_{-2.9}$ &	945$\pm$31 & $7.2^{+4.2}_{-3.4}$ \\
	     & B:  &  42.1 &	0.0$\pm$0.7 &		931$\pm$30 & $2.0^{+2.6}_{-1.6}$ \\

$\kz K^-$ & A:  &  13.8 &	0.0$\pm$1.5 &		371$\pm$19 &	$-$ \\
	     & B:  &  13.0 &	0.0$\pm$0.6 &		369$\pm$19 &	$-$ \\

$\kz\kzb$ & A:  &  8.1  &	0.0$\pm$0.5 &		34$\pm$6 &	$-$ \\
	     & B:  &  8.0  &	0.0$\pm$0.5 &		37$\pm$6 &	$-$ \\
\hline
$p\bar p$& A:  &  31.5 &	0.0$\pm$0.7 &		38$\pm$6 & $-$ \\
	     & B:  &  34.3 &$0.0^{+0.5}_{-0.6}$ &	18$\pm$5& $-$ \\

$\pL$    & A:  &  21.9 &$0.2^{+1.5}_{-0.8}$ &   46$\pm$7 & $-$ \\
	     & B:  &  21.6 &	0.0$\pm$0.8 &		44$\pm$7 &	$-$ \\

$\LL$    & A:  &  14.3 &	0.0$\pm$0.7 &		25$\pm$5 &	$-$ \\
	     & B:  &  13.4 &	0.0$\pm$0.6 &		29$\pm$6 &	$-$ \\

\hline
\end{tabular}
\label{tab:results}
\end{center}
\end{table}

%
\section{Systematic Uncertainties}\label{sec:syst}
%
The net uncertainty in our branching ratio determinations
is dominated by the statistical errors in the event yields but also
includes a systematic contribution.

We categorize systematic uncertainties in two groups, multiplicative
and additive.  Additive uncertainties are those that affect the overall
yield of signal events, while multiplicative are those that enter as
scale factors in converting the yield to a branching ratio. In view of
the following equation,
\begin{equation}
\branch(B\to X) = \frac{N_X^{\rm observed}}{\nbb \times 
{\rm (eff)} \times{\rm (secondary~BR)}}
\label{eq:BR}
\end{equation}
the multiplicative uncertainties correspond to the uncertainty in our
knowledge of the absolute number of \bbbar\ pairs in the data sample,
denoted \nbb, and the reconstruction efficiency of each mode. In
practice the uncertainties in the secondary branching ratios of
$\pi^0\to\gamma\gamma$, $\kz\to \ks\to\pi^+\pi^-$, $\Lambda\to p\pi$ and
$D^0\to (K\pi, K\pi\pz, K\pi\pi\pi)$ are negligibly small compared to
uncertainties in \nbb\ and reconstruction efficiency.

%
\subsection{Additive Systematic Uncertainties}
%

The accuracy of the signal yield obtained from the likelihood fit
depends primarily on the fidelity of the \pdfs\ used in the fit. A
secondary consideration is the correctness of the product form assumed
in Eq. \ref{eq:evtprob}, which ignores any correlations among the four
fit variable distributions. Such correlations however are expected to be
small, and  Monte Carlo tests of the fit procedure confirm this
expectation. We therefore focus on the systematic uncertainties in
signal yield which arise from systematic uncertainties in the \pdf\
parametrizations already noted in Section \ref{sec:pdfs}.  To evaluate
these uncertainties we refit the data multiple times with one \pdf\
parameter varied each time. The resulting signal variations are summed
in quadrature, separately for negative and positive yield variations, 
ignoring any correlations which may exist among the
parameters.  A representative set of these uncertainties are
displayed in Table \ref{tab:syst} for the \Kpi\ mode; details will vary from
mode to mode. (For simplicity of presentation we have combined results
from Set A and Set B, and merged the three component terms of the Fisher
\pdf.) The essential feature, however, is that the net additive systematic
error corresponds to a relative error of 3.5\% which is substantially
smaller than that statistical error, and also smaller than the
multiplicative systematic errors to be discussed next. This pattern
holds true for all modes.

\begin{table}
\begin{center}
\caption{Additive systematic errors due to \pdf\ variations for $B\to K^+\pi^-$.
Entries show change in efficiency-corrected signal yield (events) 
resulting from a parameter variation of one standard deviation up (high) or down (low).
$L$ and $R$ refer to left and right sides of an asymmetric Gaussian distribution.}
\smallskip
\begin{tabular}{|c|l|c|c|c|}
\hline
\multicolumn{3}{|c|}{Parameter}&\multicolumn{2}{c|}{Result of Parameter Variation}\cr
\multicolumn{3}{|c|}{~}& Low-variation & High-variation \cr
\hline
\multirow{4}{5mm}{\rotatebox{90}{~~~~~~~~~Signal~~~~~~~~}}&{\mb}&mean     &$-0.1$&$-0.1$\cr
                                                &               & width &$-1.4$ &$+1.3$ \cr
                                                &{\de}          &mean   &$-1.3$ &$+1.2$ \cr
                                                &               & width &$-2.8$ &$+2.5$ \cr
                                                &{\fisher}      &mean   &$-1.0$ &$+1.0$ \cr
                                                &               & width (L) &$-0.3$ &$+0.3$ \cr
                                                &               & width (R) &$-0.8$ &$+0.8$ \cr
\hline
\multirow{4}{5mm}{\rotatebox{90}{~~~~~Background~~~}}&{\mb}&$\lambda$   &$-1.1$&$+1.1$\cr
                                                &{\de}&slope     &$-0.1$&$+0.1$\cr
                                                &{\fisher}&mean  &$-0.6$&$+0.6$\cr
                                                &                & width (L) &$-0.6$ &$+0.6$ \cr
                                                &                & width (R) &$-1.0$ &$+1.0$ \cr
                                                &                & areas &$-0.8$ &$+0.8$ \cr
\hline
    & \multicolumn{2}{c|}{Total}                &    $-4.1$      &  $+3.8$\cr
\hline
\end{tabular}
\label{tab:syst}
\end{center}
\end{table}

%
\subsection{Multiplicative Systematic Uncertainties}
%
We summarize the multiplicative systematics in Table \ref{tab:mulsyst}.

The absolute number of \bbbar\ pairs in the data sample sets the scale
for all branching ratios.  We determine this number by three different
methods: counting decays of the type $B\to D^0\pi^-$, fitting
distributions of the Fox-Wolfram\cite{fox} event shape variable \rfw,
and direct computation from the run-by-run integrated luminosities, beam
energies, and the shape of the \upsi\ resonance (normalized to
1.07 nb at the peak).  The \rfw\ method was used in previous CLEO II
publications\cite{previous}, and in principle has excellent statistical
power and small systematic uncertainties, but requires substantial off-resonance
data that was not available in the first 30\% of the CLEO III running
period.  Where off-resonance data is available, the $D\pi$ method and
the \rfw\ method agree very well, and since the $D\pi$ method is
available for all data sets we use it. The direct computation technique
is used only as a check of the other methods, and is found
to be in good agreement with them. In the $D\pi$ method, three
secondary modes are used, $D^0\to K^-\pi^+$, $D^0\to K^-\pi^+\pi^0$, and
$D^0\to K^-\pi^+\pi^-\pi^+$, and a small cross-feed from $B\to DK$ is
subtracted.
  
To avoid the additional uncertainties
implied by secondary $D^0$ branching ratios, we employ CLEO II \nbb\
determinations to set the absolute scale for CLEO III:
\begin{equation}
\frac{{\RR(\bbbar)}_{III}}{{\RR(\bbbar)}_{II}}=
\frac{\RR(D\pi)_{III}}{\RR(D\pi)_{II}} 
\frac{{\epsilon}_{II}}{{\epsilon}_{III}}.
\label{eq:nbb}
\end{equation}
Event rates per unit luminosity ($\RR$) and efficiencies ($\epsilon$) are
determined separately for CLEO II (subscript $II$) and CLEO III
(subscript $III$), and for each of the three secondary decay modes. 
In the end the dominant limiting uncertainty in this technique is
the statistical error in $D\pi$ yields.

Rare \bmeson\ decay modes involving \pz, \ks, or $\Lambda$ in the final
state have additional uncertainties associated with the efficiency to
reconstruct these particles.  We determine the reconstruction efficiencies
in Monte Carlo (\effmc) simulation and then perform a separate determination in
data (\effdata).  The total error in the ratio $\effdata/\effmc$, which
includes both statistical errors and some systematic errors (such as
branching ratios) is then interpreted as the systematic uncertainty in the
reconstruction efficiency. For \pz\ the data determination consists of
measuring the ratio
\begin{equation}
\effdata(\pz) \equiv 
\frac{N(D^0\to K^-\pi^+\pi^0)/\branch(K^-\pi^+\pi^0)}{N(D^0\to K^-\pi^+)/\branch(K^-\pi^+)}
\label{eq:pzeff}
\end{equation}
where we take the ratio of $D^0$ branching ratios obtained from
reference \cite{pdg} to be $3.44\pm 0.22$.  We find $\effdata/\effmc =
1.00 \pm 0.08 \pm 0.02$ where the second error reflects conservative
uncertainty in the Dalitz amplitudes of $D^0\to K^-\pi^+\pi^0$. A
similar study was done using $\eta\to\gamma\gamma$, $\eta \rightarrow \pi^0\pi^0\pi^0$, and $\eta
\rightarrow \pi^+\pi^-\pi^0$ decays. We anticipate that further study
will refine the \pz\ systematic error estimates. A more precise
determination of the systematic error, however,  is not called for by this
analysis as any uncertainty under $\sim 20\%$ changes our
signal sensitivities only marginally. \ks\ reconstruction uncertainty is
determined similarly from comparing $D^+\to \ks\pi^+$ and $D^+\to
K^-\pi^+\pi^+$, which yields $\effdata/\effmc = 1.01 \pm 0.07$. In the
case of $\Lambda$ the comparison is of $\Lambda_c\to \Lambda \pi$ and
$\Lambda_c\to p K\pi$, and we obtain $\effdata/\effmc = 0.93 \pm 0.17$.
In the $\Lambda$ case, the net uncertainty is dominated by the
relatively poorly known branching ratios.  In all cases the systematic
uncertainties estimated by this technique are conservative (large) but
still do not dominate the final total error.

\begin{table}
\begin{center}
\caption{Multiplicative systematic errors. Entries show the fractional
change in branching ratios for each contributing source. Entries above
the line affect all modes while those below only affect modes involving
the corresponding particles. All values quoted are for CLEO III.}
\smallskip
\begin{tabular}{lc}
\hline
Source of uncertainty                         &  $\Delta\branch/\branch$ (\%) \cr
\hline\hline
Absolute number of \bbbar\ pairs              & 8\% \cr
Monte Carlo Statistics                        & 1\% \cr
\hline
Single Track Reconstruction Efficiency        & 1\% \cr
Particle ID Efficiency per Identified Track   & 3\% \cr
Single $\pi^0$ Reconstruction Efficiency      & 10\% \cr
Single $\ks$ Reconstruction Efficiency        & 7\% \cr
Single $\Lambda$ Reconstruction Efficiency    & 17\%\cr
\hline
\end{tabular}
\label{tab:mulsyst}
\end{center}
\end{table}

%
\section{CLEO III Results}
%
Event yields for the CLEO III data subsets A and B are given above in
table \ref{tab:results}.  Because the signal efficiencies of Set A and
Set B differ slightly the event yields in the two datasets do not have
exactly the same meaning and are not directly comparable or summable. To
obtain overall CLEO III results we express the measurements of Set A and
Set B in the common language of branching ratios, $\branch =
n_{sig}/({\nbb}\epsilon)$ forming the joint
likelihood $\like_{\rm stat}(\branch) = \like_A(\branch) \like_B
(\branch)$.  The subscript ``stat" emphasizes that this version of the
likelihood function reflects only statistical features of the data.  We
fold in systematic errors, which are common to Set A and Set B, by
convolving the normalized statistical likelihood function
$$
{\widehat{\like}}_{\rm stat}(\branch) = 
\frac{\like_{\rm stat}(\branch)}{\int_0^\infty \like_{\rm stat}(\beta)d\beta} 
$$ 
with both additive event
yield uncertainties $\eta$, distributed according to an asymmetric
Gaussian, $\GG(\eta)$, and multiplicative scale factor uncertainties
$\rho$, distributed according to a symmetric Gaussian $G(\rho)$. The
widths of these distributions have been discussed above. Formally this
convolution may be written
\begin{equation}
\widehat{\like}(\branch)=
\int_{-\infty}^{\infty}d\rho
\int_{-\infty}^{\infty}d\eta
~{\widehat{\like}}_{\rm stat}
\left(\frac{\nbb\branch\epsilon+\eta}{\nbb\epsilon(1+\rho)}\right) 
\GG(\eta)G(\rho).
\label{eq:convolution}
\end{equation}
\noindent For convenience the double convolution is performed
by a Monte Carlo method.

The resulting distribution of ${\widehat{\like}}(\branch)$ is the final CLEO III likelihood
function including all of the uncertainties in the measurement.  From it
we find the minimum of the $-2 \ln {\cal L}$ distribution to measure our
mean, and find the 1$\sigma$ intersections to determine the errors. 
Since this is the total error, we unfold the systematic error by
subtracting the statistical error in quadrature from the total error. We
set 90\% confidence level upper limits by determining the value of
\branch\ for which 
$$
\int_0^{\branch}\widehat{\like}(\beta)d\beta=0.90,
$$
and calculate
significances by looking at the zero yield value of the $-2 \ln {\cal
L}$ distribution. In the limit of a purely Gaussian likelihood function,
this definition of significance reduces to the signal yield
divided by its one standard deviation error.

%
\section{Combined CLEO II and CLEO III Results}
%
We combine CLEO II and CLEO III measurements using the likelihood
functions described above for CLEO III and reported in Ref.
\cite{lastkpipaper} for CLEO II. For some modes we use previously
unpublished likelihood functions. The baryonic modes and the \ks\ks\
mode were analyzed here with the full CLEO II data set for the first
time. 

\begin{figure*}[htbp]
\begin{center}
\mbox{
\includegraphics*[width=6.0in]{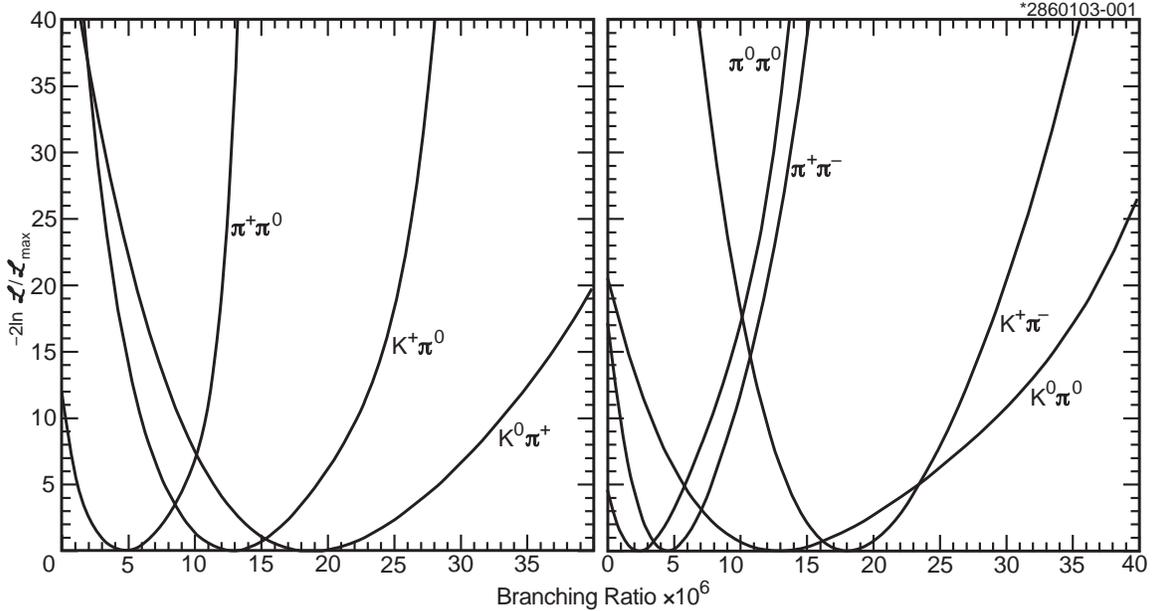}  
}
\end{center}
\caption{$-2 \ln ({\cal L}/{\cal L}_{max})$ distributions for CLEO II and CLEO III combined,
for $K\pi$ and $\pi\pi$ modes with non-zero yield.} 
\label{fig:pipi-all}
\end{figure*}

\begin{figure}[htbp]
\begin{center}
\includegraphics*[width=3.0in]{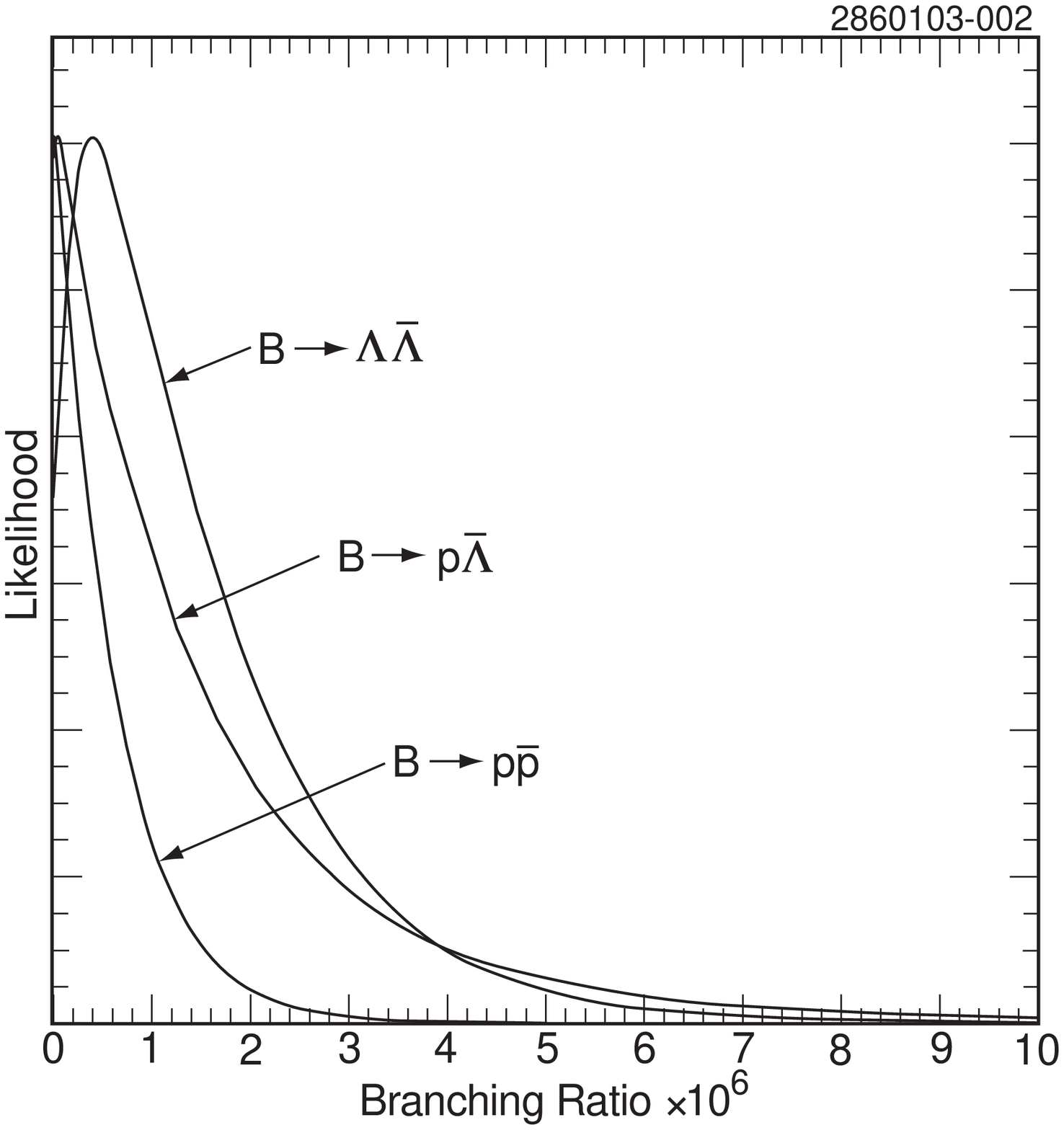}  
\end{center}
\caption{Likelihood functions for $B\to\ppbar$, $B\to \pL$, and $B\to\LL$.} 
\label{fig:baryons}
\end{figure}

Particle identification in CLEO II was limited, and modes with potential
for $K/\pi$ misidentification, such as $B\to \pi^+\pi^-$ and $B\to
K^+\pi^-$, were analyzed in terms of two-dimensional likelihood
functions, $\like(N_{\pi^+\pi^-}, N_{K^+\pi^-})$.  The improved $K/\pi$
separation in CLEO III however permits us to treat these modes
independently in the new data. To combine CLEO II and CLEO III
likelihood functions, therefore, we first project the two-dimensional
CLEO II functions on to one-dimensional versions, using $\like(x) = \int
\like(x,y) dy$, and then express in terms of branching ratios,
$\like(\branch)=\like(N_{sig}/({\nbb}\epsilon))$. Systematic errors are
included following the same method as described for CLEO III above to
obtain a total CLEO II likelihood function for each mode.  A final
combined CLEO II and CLEO III likelihood function is then formed from
the joint likelihood, $\like_{\rm final}(\branch) = \like_{\rm CLEO
II}(\branch)\like_{\rm CLEO III}(\branch)$. For $\pi\pi$ and $K\pi$
modes with non-zero yields we plot the negative log-likelihood functions
in Fig. \ref{fig:pipi-all}. Likelihood functions for the
di-baryonic modes are shown in Fig. \ref{fig:baryons}.  Table
\ref{table:cleoAll} summarizes the final results, with separate entries
for CLEO II results (extracted from the references and reproduced here
for the convenience of the reader), CLEO III results, and the combined
CLEO II and CLEO III results.

\begin{table*}[htbp]
\begin{center}
\caption{Experimental results for CLEO II, CLEO III, and
both datasets combined.  Significances include systematic errors.
Note that the $p{\bar p}$ analysis in Ref. 4 was 
done in only a subset of the full CLEO II dataset, so the 
``combined'' result is simply the CLEO III upper limit.  Upper limits
are 90\% confidence level. CLEO II results are taken from Ref. 4,
except for the $\kz\kzb$, \pL\, and \LL\ modes which were analyzed
in this work with the full CLEO II dataset for the first time.}
\smallskip
\begin{tabular}{|l|cc|cc|cc|}
\hline
 & \multicolumn{2}{|c|}{CLEO II - Ref. 4}
 & \multicolumn{2}{|c|}{CLEO III} 
 & \multicolumn{2}{|c|}{Combined}\\
\hline
Mode & Significance & $\branch\power{6}$
     & Significance & $\branch\power{6}$
     & Significance & $\branch\power{6}$ \\ \hline\hline
$\pi^+\pi^-$ & 4.2    & 4.3$^{+1.6+0.5}_{-1.4-0.5}$ 
             & 2.6    & 4.8$^{+2.5+0.8}_{-2.2-0.5}$ 
             & 4.4    & 4.5$^{+1.4+0.5}_{-1.2-0.4}$ \\
$\pi^+\pi^0$ & 3.2    & 5.6$^{+2.6+1.7}_{-2.3-1.7}$  
             & 2.1    & 3.4$^{+2.8+0.8}_{-2.0-0.3}$ 
             & 3.5    & 4.6$^{+1.8+0.6}_{-1.6-0.7}$ \\
$\pi^0\pi^0$ & 2.0    & $(<5.7)$                    
             & 1.8    & $(<7.6)$                    
             & 2.5    & $(<4.4)$                    \\ \hline
$K^+\pi^-$   & $12$   & 17.2$^{+2.5+1.2}_{-2.4-1.2}$
             & $ >7$  & 19.5$^{+3.5+2.5}_{-3.7-1.6}$
             & $>7$   & 18.0$^{+2.3+1.2}_{-2.1-0.9}$\\
$\kz\pi^+$   & 7.6    & 18.2$^{+4.6+1.6}_{-4.0-1.6}$
             & 4.6    & 20.5$^{+7.1+3.0}_{-5.9-2.1}$
             & $>7$   & 18.8$^{+3.7+2.1}_{-3.3-1.8}$\\
$K^+\pi^0$   & $6.1$  & 11.6$^{+3.0+1.4}_{-2.7-1.3}$
             & $5.0$  & 13.5$^{+4.0+2.4}_{-3.5-1.5}$
             & $>7$   & 12.9$^{+2.4+1.2}_{-2.2-1.1}$\\
$\kz\pi^0$   & 4.9    & 14.6$^{+5.9+2.4}_{-5.1-3.3}$  
             & 3.8    & 11.0$^{+6.1}_{-4.6}\pm2.5$  
             & 5.0    & 12.8$^{+4.0+1.7}_{-3.3-1.4}$\\ \hline
$K^+K^-$     & -      & $(<1.9)$                    
             & -      & $(<3.0)$                    
             & -      & $(<0.8)$                    \\
$\kz K^-$    & -      & $(<5.1)$                    
             & -      & $(<5.0)$                    
             & -      & $(<3.3)$                    \\
$\kz\kzb$    & -      & $(<6.1)$                    
             & -      & $(<5.2)$                    
             & -      & $(<3.3)$                    \\ \hline
$\ppbar$     & -      & $(<7.0)$                    
             & -      & $(<1.4)$                    
             & -      & $(<1.4)$                    \\
$\pL$        & -      & $(<2.0)$                    
             & -      & $(<3.2)$                    
             & -      & $(<1.5)$                    \\
$\LL$        & -      & $(<1.8)$                    
             & -      & $(<4.2)$                    
             & -      & $(<1.2)$                    \\ \hline
\end{tabular}
\label{table:cleoAll}
\end{center}
\end{table*}

%
\section{Physically Interesting Ratios and the Phase of $V_{ub}$}
%

As discussed in the introduction, it is possible to extract information
about the phase of \vub\ from these charmless \bmeson\ decay data. The
method of Reference 3 is based on two ratios of the branching fractions
which we have measured and reported above.  Using the notation of this
reference, and combining statistical and systematic errors
in quadrature, the ratios are found to be:

\begin{equation}
R_{*}\equiv
\frac{\branch(B^\pm \to K^0 \pi^\pm)}{2\branch(B^\pm \to K^\pm\pi^0)}
= 0.73 \pm 0.21,
\end{equation}

and

\begin{equation}
\epsilon_{exp}\equiv \tan\theta_C\frac{f_K}{f_\pi}
\left[
\frac{2\branch(B^\pm \to \pi^\pm\pi^0)}{\branch(B^\pm \to K^0 \pi^\pm)}
\right]^{1/2} = 0.18 \pm 0.04.
\end{equation}

\begin{figure}[hbtp]
\begin{center}
\mbox{
\includegraphics*[width=3.0in]{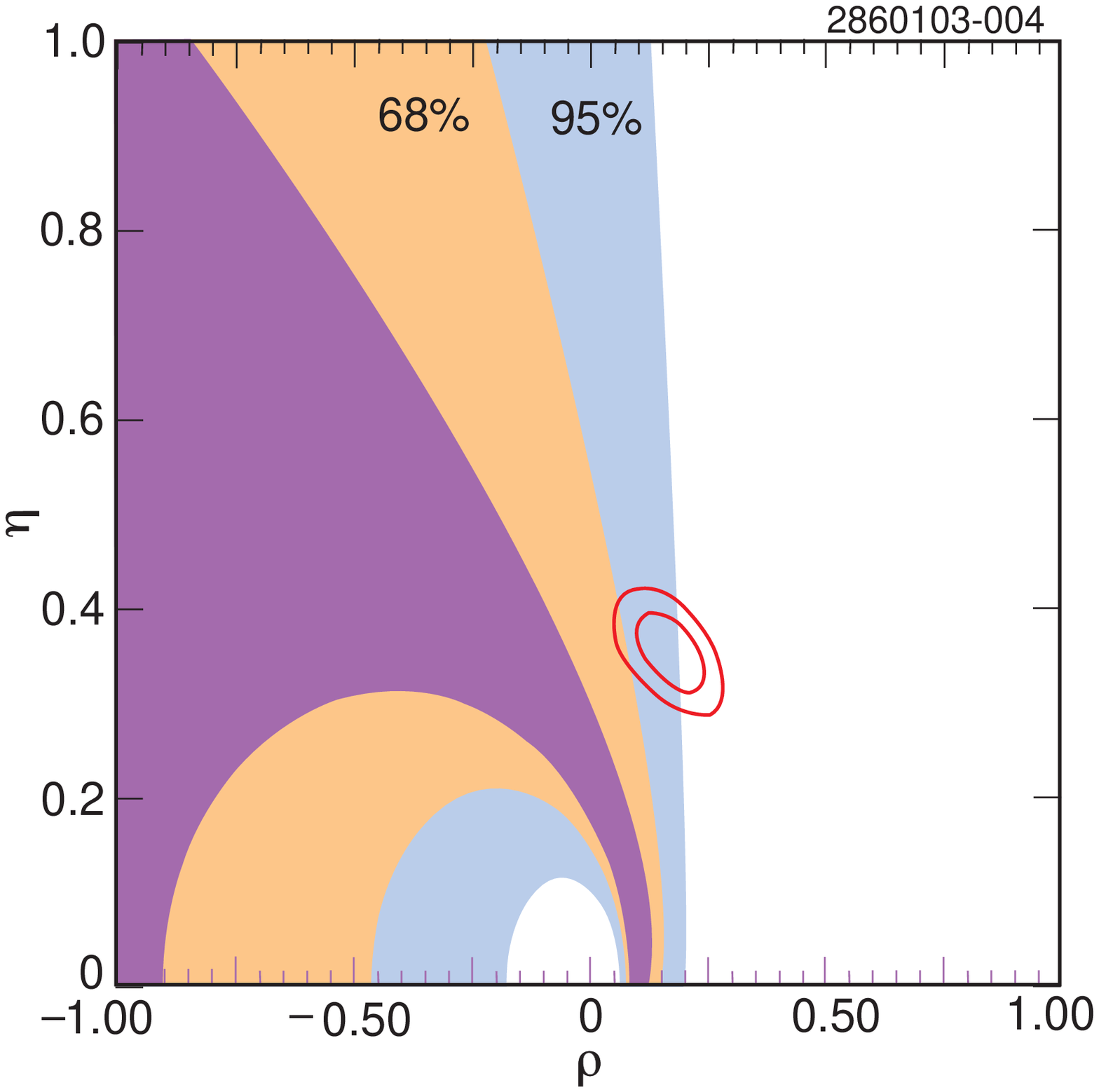}  
}
\end{center}
\caption{Confidence contours in the $\eta-\rho$ plane.  The shaded
bands represent regions allowed by the world-averaged charmless \bmeson\ decay
measurements while the ellipsoids represent 68\% and 95\% contours
from conventional global fits to heavy quark measurements. The
dark shaded region corresponds to the experimental central values
of the charmless data, smeared by theoretical uncertainty.  See
text for details and references.}
\label{fig:alan}
\end{figure}

We see that the precision available with the CLEO data is about 20-30\%
in these quantities. With data from the {\it BABAR} and Belle
experiments\cite{babarandbelle} we can make world (weighted) averages of
branching ratios and reach 10-15\% experimental precision in the
critical ratios: $R_{*}=0.71\pm 0.09$ and $\epsilon_{exp}=0.21\pm 0.02$.
 These numbers in turn indicate a preferred region for $\gamma =
Arg(\vub)$ which is greater than 90${}^\circ$\cite{alan}. Using these
world-averaged data we construct contours in the $\rho-\eta$ plane
according to prescription of Reference 3 and display the result in
Figure \ref{fig:alan}.  The dark band represents the experimental
central value convolved with theoretical uncertainties; lighter bands
show the additional coverage when 68\% and 95\% experimental confidence
regions are included. For reference we also overlay  68\% and 95\%
confidence level ellipses of the preferred apex of the unitarity
triangle as obtained in a standard analysis based on $B$ mixing, $\sin
2\beta$, \vub, and kaon decays\cite{stochhi}. An intriguing discrepancy
between these regions is noticeable.  In the short term the most
substantial progress to be made will be in reducing the statistical
errors on the branching ratios of charmless \bmeson\ decay modes. If
discrepancies survive there could be non-trivial implications for the
Standard Model, as discussed in Ref. 3.

%
\section{The $\branch(B\to DK)/\branch(B\to D\pi)$ ratio}
%

\begin{figure*}[hbtp]
\begin{center}
\mbox{
\includegraphics*[width=6.0in]{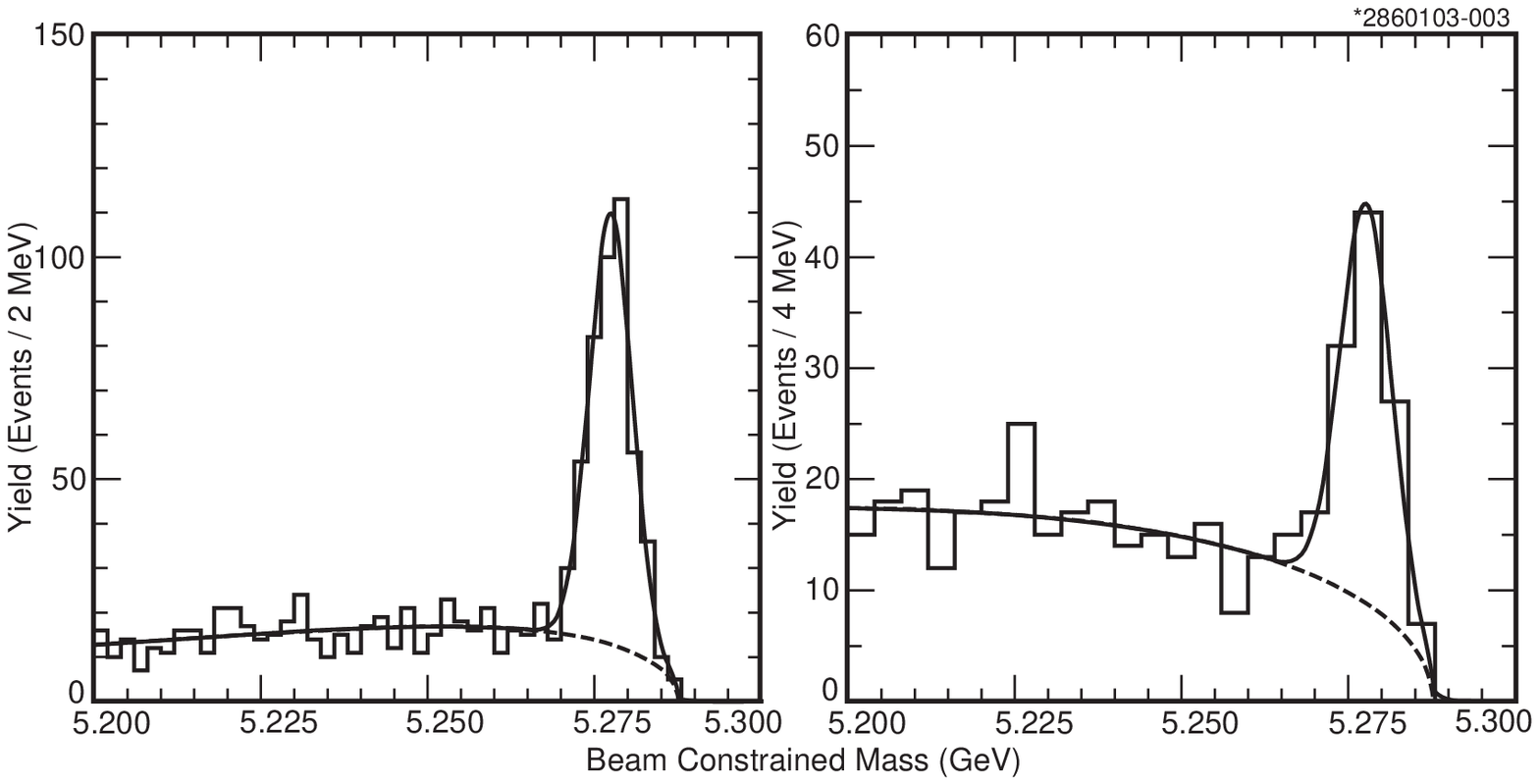} 
}
\end{center}
\caption{CLEO III data: the $M_{B}$ distribution for  $B^- \rightarrow D^0\pi^-$ 
(left) and $B^- \rightarrow D^0K^-$ (right) candidates.}
\label{fig:kamal}
\end{figure*}

In view of the good $K/\pi$ separation in CLEO III data we also report a
new determination of the ratio $\branch(B^-\to D^0K^-)/\branch(B^-\to
D^0\pi^-)$ which benefits substantially from good particle
identification. The original CLEO II publication is available
in Ref. \cite{abi}.

For this analysis, $D^0$ candidates are reconstructed in three secondary
modes, $D^0\to K^-\pi^+$, $D^0\to K^-\pi^+\pi^0$, and $D^0\to
K^-\pi^+\pi^-\pi^+$. Requirements for the $B\to D^0h^-$ modes include a
30 MeV $D^0$ mass cut, a 100 MeV \de\ cut, and standard particle ID as
described previously on both the primary $h^-$ from the \bmeson\ and on
the secondary kaon from the $D^0$.  The $\pi^0$ mass for the $D^0 \ra
K^-\pi^+\pi^0$ mode is required to be within 30 MeV of its nominal
mass. For the $D^0 K$ likelihood fit, $D^0 \pi$ is included as a
cross-feed background, and  corresponds to 
approximately 50\% of the DK yield shown in Fig. \ref{fig:kamal}.
(Both $D^* \pi$ and $D\rho$ were found not to be
significant backgrounds to either signal mode.) Fig. \ref{fig:kamal}
shows \mb\ distributions for $B^-\to D^0\pi^-$ and $B^-\to D^0 K^-$ candidates
with the likelihood fit shape superimposed.

Combining the three $D^0$ submodes, we find
\begin{equation}
\frac{\branch(B^-\ra D^0 K^-)}{\branch(B^-\ra D^0 \pi^-)} = 
(9.9^{+1.4+0.7}_{-1.2-0.6})\power{-2}.
\end{equation}
Most systematic errors cancel in this ratio, with only a small
residual arising from the particle identification requirements
imposed on the primary $\pi/K$ in both numerator and denominator.

%
\section{Summary}
%
We have presented final results from the CLEO experiment on charmless hadronic
\bmeson\ decays. The decay modes include the ten \pipi, \Kpi, and \kk\
final states as well as the dibaryonic states $p\bar p$, $p\bar \Lambda$
and \LL.  In addition we have presented a new determination of the ratio
of branching ratios $\branch(B\to DK)/\branch(B\to D\pi)$. The results
are based on the full CLEO II and CLEO III data samples totalling
15.3 \fbinv\ at the $\Upsilon(4S)$, and supercede previously
published results by this collaboration.

%
\section{Acknowledgements}
%
We gratefully acknowledge the effort of the CESR staff in providing us
with excellent luminosity and running conditions. M. Selen thanks the
Research Corporation, and A.H. Mahmood thanks the Texas Advanced
Research Program. This work was supported by the National Science
Foundation, and the U.S. Department of Energy.

%
%

%
%
\def\endpoint{;~~}
\def\Journal#1&#2&#3(#4){#1{\bf #2}, #3 (#4)}
\def\NIM{Nucl. Instr. and Meth. }
\def\NIMA{Nucl. Instr. and Meth. A }
\def\NPB{Nucl. Phys. B }
\def\PLB{Phys. Lett. B }
\def\PRL{Phys. Rev. Lett. }
\def\PRD{Phys. Rev. D }

\end{document}